%% file: b.tex
\documentstyle[12pt,twoside,fleqn,espcrc1,epsf]{article}
\newcommand{\AmS}{{\protect\the\textfont2
  A\kern-.1667em\lower.5ex\hbox{M}\kern-.125emS}}

\title{Cosmological and astrophysical aspects of finite-density QCD}  

\author{D. J. Schwarz\address{Institut f\"ur Theoretische Physik, 
        Universit\"at Frankfurt,\\ 
        Postfach 11 19 32, 60054 Frankfurt am Main, Germany}
        \thanks{e-mail: dschwarz@th.physik.uni-frankfurt.de}}

\begin{document}
\maketitle

\begin{abstract}
The different phases of QCD at finite temperature and density
lead to interesting effects in cosmology and astrophysics.
In this work I review some aspects of the cosmological QCD transition
and of astrophysics at high baryon density.
\end{abstract}

\section{Introduction}

In the early universe, at high temperatures and almost 
vanishing chemical potential, we expect from asymptotic freedom that quarks
and gluons are deconfined in a quark-gluon plasma (QGP).
As the universe cools by expansion, a transition to the confined phase,
a hadron gas (HG), occurs at $T_\star \sim \Lambda_{\rm QCD}$. 
Lattice QCD at vanishing chemical potential provides important insight into
the nature of the cosmological QCD transition \cite{lattice}.

In the most compact stars, neutron stars or quark stars, another transition 
associated with QCD might happen. The dense phase might be a QGP at 
high chemical potential and low temperature, or it might be color 
superconducting phase \cite{Wilczek,Wtalk} of diquarks. 
Between the cosmological and the astrophysical branch of the QCD phase diagram 
is the region where heavy ion collisions \cite{Herrmann} take place.
 
In Sec.~2 I sketch the evolution of the baryon density and the 
baryon chemical potential during the expansion of the universe. 
The baryon chemical potential is almost zero in the very early
universe. After the annihilation of all anti-baryons (at $T_{\rm ann} 
\sim m_{\rm p}/25$) the baryon chemical potential is approximately equal 
to the nucleon rest mass.

The cosmological QCD transition is discussed in Sec.~3.  
I show that strangelet formation \cite{Witten}
and inhomogeneous nucleosynthesis \cite{IBBN} are unlikely consequences 
of the QCD transition. The reason is the small mean bubble separation 
(much smaller than the Hubble radius) \cite{Ignatius,Christiansen}, which 
follows from recent lattice QCD results for latent heat and surface tension
\cite{Iwasaki2}. At scales larger than the bubble separation 
the sound speed vanishes during the transition \cite{SSW,Jedamzik}. 
This means that density perturbations fall freely during the transition 
giving rise to large amplifications of the primordial density
spectrum at small 
scales \cite{SSW}. This leads to the formation of small clumps of cold dark 
matter ($M_{\rm clump} < 10^{-10} M_{\odot}$), for cold dark matter which is 
kinetically decoupled at the QCD scale, e.g.\ axions. 
The large drop in relativistic degrees of freedom during the QCD transition
modifies the spectrum of primordial gravitational waves \cite{Schwarz}.  

High baryon densities are realized in compact astrophysical bodies like 
neutron, hybrid, or (strange) quark stars. Lattice QCD cannot deal with 
a finite chemical potential so far \cite{Barbour}, therefore the 
equation of state of dense matter is unknown. Toy models, like the bag model 
are used instead. Observable consequences of the 
high density phase(s) might be visible in the cooling curves of neutron stars
\cite{Pethick,Schaab}. A phase transition inside a pulsar might be observed 
as severe spin-up of the pulsar, as recently shown by 
Glendenning, Pei, and Weber \cite{Glendenning}.   

Finally, I comment on the ultimate heavy 'ion' collision, the collision of 
two neutron stars. It has been suggested that such an event might be the 
central engine of short gamma-ray bursts \cite{Katz}. 

\section{Baryon number density and chemical potential}

At high temperatures ($T > \Lambda_{\rm QCD}$) the baryon number density
may be defined as $n_{\rm B} \equiv \frac 13 \sum (n_q - n_{\bar{q}})$, 
where $n_q$ ($n_{\bar{q}}$) is the number density of a specific quark 
(anti-quark) flavor, and the sum is taken over all quark flavors. At $T < 1$ 
GeV only the u, d, and s quarks contribute significantly. At low temperatures 
($T < \Lambda_{\rm QCD}$) the baryon number density is defined as $n_{\rm B} 
\equiv \sum (n_b - n_{\bar{b}})$, now the summation is taken over 
all baryon species. Practically the nucleons contribute to the baryon number 
of the universe only.

Below the electroweak transition ($T_{\rm ew} \sim 300$ GeV) the baryon 
number, $B$, in a comoving volume is conserved. On the other hand 
entropy, $S$, is conserved.
As a consequence the ratio of baryon number density and
entropy density, $s$, is constant. From the abundances of primordial 
${}^4$He and D [produced in big bang nucleosynthesis (BBN)]
we know the ratio $n_{\rm B}/n_{\gamma} = 
(1.5 \mbox{\ to\ } 6.3) \times 10^{-10}$ \cite{BBN}. 
Taking into account the three massless neutrinos along with the photons that 
contribute to the entropy density, we find  
\begin{equation}
\label{nB}
{n_{\rm B}\over s} = (2 \mbox{\ to\ } 8) \times 10^{-11} \ .
\end{equation}
Due to this very small ratio the number of quarks equals the 
number of anti-quarks in the very early universe. 

Let me now turn to the baryon chemical potential. At high temperatures 
the quark chemical potentials, $\mu_{\rm q}$, are equal, because weak 
interactions keep them in chemical equilibrium (e.g.\ u + e $\leftrightarrow$ 
d or s + $\nu_{\rm e}$), and the chemical potentials for the leptons are 
assumed to vanish (see \cite{Weinberg} for a discussion of lepton chemical 
potentials). Thus, the chemical potential for a baryon is defined by 
$\mu_{\rm B} \equiv 3 \mu_{\rm q}$. For an anti-baryon the chemical potential
is $ - \mu_{\rm B}$. The baryon number density of an ideal Fermi gas 
of three quark flavors reads $n_{\rm B} 
\approx T^2\mu_{\rm B}/3$ at high temperature $T$. From Eq.~(\ref{nB}) we 
find that
\begin{equation}
{\mu_{\rm B}\over T} \sim 10^{-9} \qquad \mbox{at\ }
T > \Lambda_{\rm QCD} \ . 
\end{equation}
Thus, lattice QCD at finite temperature and vanishing chemical potential
is most appropriate to study the QCD equation of state for the early universe. 

At low temperatures ($T< \Lambda_{\rm QCD}$) $\mu_{\rm B} = \mu_{\rm p} = 
\mu_{\rm n}$, neglecting the mass difference between the proton and the 
neutron. The ratio of baryon number density and entropy now reads
\begin{equation}
\label{mu}
{n_{\rm B}\over s} \approx 0.05 \left({m_{\rm p}\over T}\right)^{\frac{3}{2}}
\exp\left(-{m_{\rm p}\over T}\right) \sinh\left({\mu_{\rm B}\over T}\right) \ .
\end{equation}
Since this ratio is constant, the behavior of $\mu_{\rm B}/T$ is given by 
Eq.\ (\ref{mu}), e.g.\ at $m_{\rm p}/T \approx 20$ we find 
$\mu_{\rm B}/T \approx 10^{-2}$. All anti-baryons
are annihilated when the ratio
\begin{equation}
{n_{\rm b} - n_{\bar{\rm b}}\over n_{\rm b} + n_{\bar{\rm b}}} \approx
\tanh\left({\mu_{\rm B}\over T}\right) 
\end{equation} 
goes to unity. This happens when $\mu_{\rm B}/T \sim 1$, which corresponds to
$m_{\rm p}/T_{\rm ann} \approx 25$ or $T_{\rm ann} \approx 40$ MeV. 
Below this temperature the baryon chemical 
potential is $\mu_{\rm B}(T\ll T_{\rm ann}) \approx m_{\rm p}$. 
To add one proton to the universe one proton rest mass should be invested.

\section{The cosmological QCD transition}

Recent lattice QCD results for two quark flavors suggest that QCD makes
a transition at a temperature of $T_\star \sim 150$ MeV \cite{lattice,MILC96}. 
Lattice calculations with three quark flavors indicate that the transition 
is of first order for the physical mass of the strange quark 
\cite{Iwasaki}. These calculations use Wilson quarks. However, with 
staggered quarks it has been concluded that the QCD transition is a crossover 
for the physical quark masses \cite{Brown}.
I therefore discuss a first order QCD transition and a QCD crossover below.

Lattice QCD with quenched quarks shows a first order transition with
a latent heat $l = 1.4 T_\star^4$ and a surface tension 
$\sigma = 0.015 T_\star^3$ \cite{Iwasaki2}. These values are much smaller
than the values suggested by the bag model. $l$ and $\sigma$ are not 
known for the physical masses of the quarks (there is just an upper limit
$\sigma < 0.1 T^3_\star$ \cite{Hackel}). 

The scale of the cosmological QCD transition is given by the Hubble radius,
$R_{\rm H}$, at the transition, which is $R_{\rm H} \sim m_{\rm Pl}/T^2_\star
\sim 10$ km. The mass inside the Hubble volume is $\sim 1 M_\odot$.
The expansion time scale is $10^{-5}$ s, which should be compared with the
time scale of QCD, $1$ fm/c $\approx 10^{-23} s$. All interaction rates
are much higher than the expansion rate of the universe. Even 
the rate of weak interactions, $\Gamma_{\rm w} \sim G_{\rm F}^2 T^5$, exceeds
the Hubble rate, $H \sim T^2/m_{\rm Pl}$, by a factor $10^7$. Therefore
photons, leptons, quarks and gluons (or pions) are tightly coupled and may be 
described as a single, adiabatically expanding fluid. 

The cosmological QCD transition causes a dramatic drop in the number of 
relativistic degrees of freedom, $g_*$. In the quark-gluon plasma 
$g_*^{\rm QGP} = 61.75 (51.25)$ with(out) the strange quark, whereas
for the hadron gas $g_*^{\rm HG} = 17.25 (21.25)$ including pions (plus kaons 
and eta) besides the photon and leptons ($e,\mu,\nu$s). 

This large drop in degrees of freedom modifies the spectrum of primordial
gravitational waves \cite{Schwarz}. One source of primordial gravitational 
waves is inflation \cite{Starobinskii}.
These gravitational waves are produced with an almost scale invariant spectrum, 
i.e.\ the root mean square (rms) amplitude $h_f$ of frequency $f$ is the 
same for all 
frequencies at horizon crossing, $f = H$. After horizon crossing
the energy density of the gravitational waves is redshifted, i.e. $\rho_{\rm gw}
\sim m_{\rm Pl}^2 f^2 h_f^2 \sim a^{-4}$, where $a$ is the scale factor of 
the universe.
Without the QCD transition this results in a constant energy density of 
gravitational waves per frequency interval, $\Omega_{\rm gw}(f) 
\equiv ({\rm d} \rho_{\rm gw} /{\rm d} \ln f)/\rho_{\rm crit} = $ const., 
for modes that 
cross the Hubble radius before radiation-matter equality. If we now 
take the large drop of degrees of freedom during the QCD transition into 
account, 
the Hubble radius evolves as $R_{\rm H} \sim g_*^{1/6}(a) a^2$. On the 
other hand, 
physical scales grow like $a$. Thus, the number of modes that enter the Hubble 
radius per time interval changes, and therefore the slope of 
$\Omega_{\rm gw}(f)$
changes during the QCD transition. Modes that enter after the transition
are unaffected, whereas modes that enter long before the transition 
are suppressed by a factor 
\begin{equation}
{\Omega_{\rm gw}(f \gg f_\star) \over \Omega_{\rm gw}(f \ll f_\star)} =
\left(g_*^{\rm HG}\over g_*^{\rm QGP}\right)^{\frac13} \approx 
0.7 \ .
\end{equation}
This step is shown in Fig.~1. It is also demonstrated that this step
in the spectrum is the same for a first order phase transition (bag equation 
of state) and for a crossover (interpolation between both phases by a tanh). 
The chances to detect this step in pulsar timing residuals \cite{pulsar} in the
near future is poor, because the expected rms amplitude of primordial 
gravitational waves is orders of magnitudes below the current pulsar timing 
sensitivity. However, the only observable consequence of a QCD crossover is,
to my knowledge, the step in the spectrum of gravitational waves. 
\begin{figure}
\begin{center}
\input{om.tex}
\vspace{- 1cm}
\end{center}
\caption{The modification of the energy spectrum, per logarithmic frequency
interval, for primordial gravitational waves from the QCD transition 
\cite{Schwarz}.
\label{fig1}}
\vspace{- \baselineskip}
\end{figure}
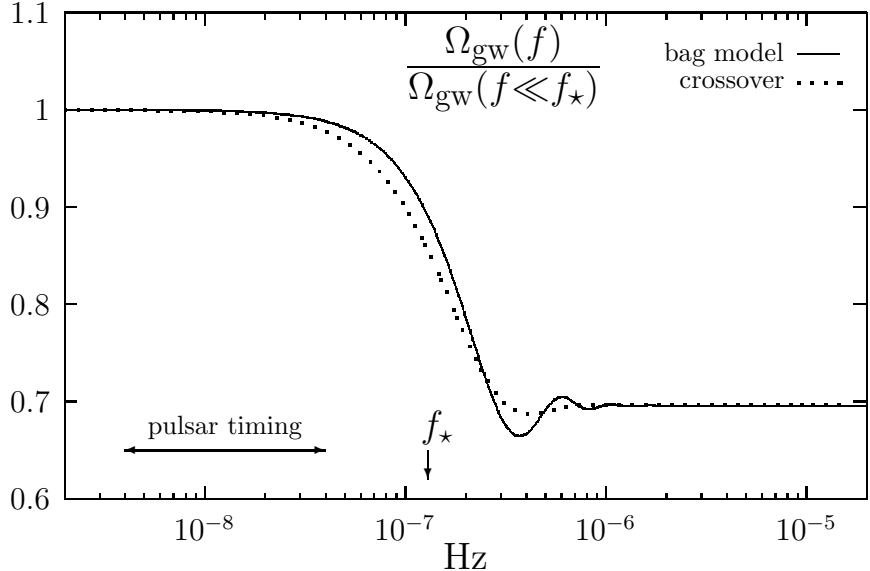

\subsection{Effects from the bubble separation scale}

For a first order QCD transition hadronic bubbles nucleate during a short
period of supercooling, $\Delta t_{\rm sc} \sim 10^{-8}$ s \cite{Ignatius}.
The typical bubble nucleation distance is a few cm for homogeneous
nucleation \cite{Christiansen}, which is $d_{\rm nucl} \approx 
10^{-6} R_{\rm H}$. The hadronic bubbles grow very fast, until the 
released latent heat has reheated the universe to $T_\star$. This happens 
almost instantaneously. The increase in entropy due to this reheating is just 
$\Delta S \sim 10^{-6} S$. For the remaining $99.9\%$ of the
transition the HG and the QGP coexist at the pressure 
\begin{equation}
p_\star \equiv p_{\rm HG}(T_\star) = p_{\rm QGP}(T_\star) \ .
\end{equation}
During this time the hadronic bubbles grow slowly
and the release of latent heat keeps the temperature constant until the
transition is completed. At the end of the transition only few quark droplets
are left over, with a typical separation $d_{\rm nucl}$.

In the mid 80s interest in the cosmological QCD transition arose, because
it was realized that a strong first order QCD phase transition could lead to
important observable signatures. Witten \cite{Witten} pointed out that a
separation of phases during the coexistence of the hadronic and the
quark phase could gather most baryons in (strange) quark nuggets 
\cite{sm,Witten}. These would contribute to the dark matter today. 
At the end of the transition the baryon number in the quark droplets 
could exceed the baryon number in the hadron phase by several orders of 
magnitude, $n_{\rm B}^{\rm QGP}$ could be close to nuclear density 
\cite{Sumiyoshi}. However, it was realized
that the quark nuggets, while cooling, lose baryons. The quark nuggets
evaporate, unless they contain much more than $10^{44}$ baryons initially
\cite{evap}. This number should be compared with the number of baryons
inside a Hubble volume at the QCD transition, which is $10^{50}$.
Thus, the mean bubble nucleation distance should be $> 10^{-2} R_{\rm H}
\sim 100$ m in order to collect enough baryons.
 
In \cite{Sumiyoshi,evap} a chromoelectric flux tube model
was used to estimate the penetration rate of baryons through the interface.
A quark that tries to penetrate the interface creates a flux tube, which 
most probably breaks up into a quark anti-quark pair. By this mechanism
mesons evaporate easily. On the other hand baryons are formed rarely, 
because a diquark anti-diquark pair has to be produced in the break up
of the flux tube. 
One could easily think of mechanisms that would increase the evaporation
rate of baryons \cite{Hofmann}.
If a significant fraction of diquarks was formed in the quark phase, 
these diquarks could penetrate the interface by creating a flux tube, which
eventually breaks creating a quark anti-quark pair. The quark would leave 
together with the diquark and form a baryon, whereas the anti-diquark 
would remain 
in the quark phase. Such a mechanism would increase the evaporation rate 
dramatically. In this case even unrealistically large values of $d_{\rm nucl}$
would not give rise to the formation of strangelets.

Applegate and Hogan found that a strong first order QCD phase transition
induces inhomogeneous nucleosynthesis \cite{BBN}. It is extremely 
important to understand the initial conditions for BBN, because many of 
our ideas about the early universe rely on the validity of the standard
(homogeneous) BBN scenario. The standard BBN scenario is in 
good agreement with observations \cite{BBN}. In inhomogeneous nucleosynthesis
large isothermal fluctuations of the baryon number (the remnants of the
quark droplets at the end of the QCD transition) could lead to different 
yields of light elements. As a minimal
requirement for an inhomogeneous scenario of nucleosynthesis
the mean bubble nucleation distance
has to be larger than the proton diffusion length, which corresponds
to $\sim 3$ m \cite{Mathews} at the QCD transition. This is two orders
of magnitude above recent estimates of the typical nucleation distance 
\cite{Christiansen}.
On the other hand the observed cosmic abundances of light elements
do not favor inhomogeneous nucleosynthesis, except a small region in
parameter space corresponding to an inhomogeneity scale of $\sim
40$ m \cite{Mathews}. 

The above effects deal with physics at the bubble separation scale.
It was assumed that the mean bubble separation is large ($\sim 100$ m),
because latent heat and surface tension had been
overestimated in the past. Recent lattice QCD calculations \cite{Iwasaki2} 
find smaller values for latent heat and surface tension and therefore
the mean bubble nucleation distance is just a few cm. 

\subsection{Effects from the Hubble scale}

\begin{figure}[t]
\begin{center}
\epsfig{figure=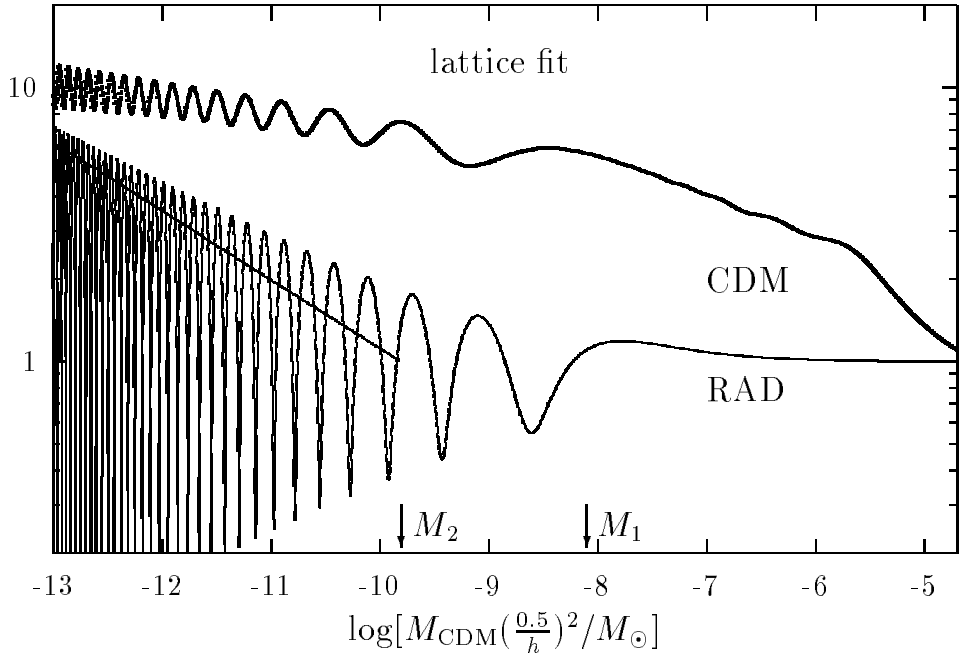,width=0.8\linewidth}
\end{center}
\vspace{-1.3 cm}
\caption{
The modifications of the density contrast, $\delta \equiv 
\delta\rho/\rho$, of kinetically decoupled CDM (like axions) 
and of the radiation fluid fluctuation amplitude (RAD) due to the 
QCD transition as obtained in \cite{SSW}. Both 
quantities are normalized to the pure Harrison-Zel'dovich radiation 
amplitude. On the horizontal axis the wavenumber $k$ is represented by the 
CDM mass contained in a sphere of radius $\pi/k$. The QCD equation of state 
has been fitted to lattice QCD results \cite{Boyd,MILC97}. The 
straight line starts at $M_2$ and shows the asymptotic envelope for 
small scales $\propto k^{3/4}$. The CDM mass inside the Hubble 
radius is denoted $M_1$.}
\vspace{- \baselineskip}
\end{figure}
Scales $\lambda$ that are of the order of the Hubble radius $R_{\rm H}$ are 
not sensitive to details of the bubbles. Together with Schmid and Widerin
I found in Ref.\ \cite{SSW} that the evolution of
cosmological density perturbations is strongly affected by a first order QCD
transition for subhorizon scales, $\lambda < R_{\rm H}$. Cosmological
perturbations on all scales are predicted by inflation
\cite{perturbations} and have been observed in the temperature
fluctuations of the cosmic microwave background by the COBE satellite
\cite{Smoot}.

In the radiation dominated universe subhorizon density perturbations
perform acoustic oscillations. The restoring force is provided by
pressure gradients. The pressure gradients, hence the isentropic sound 
speed $c_s = \left(\partial p/\partial \rho \right)_S^{1/2}$
(on scales much larger than the bubble separation scale) drops to
zero at a first order QCD transition \cite{SSW}, because both phases
coexist at the pressure $p_\star$ only ($a$ is the scale factor of 
the universe):
\begin{equation}
c_s^2 \equiv \left(\partial p \over \partial \rho \right)_S = 
{{\rm d} p_\star/{\rm d} a\over {\rm d} \rho(a)/{\rm d} a} = 0 \ .
\end{equation}
It stays zero during the 
entire
transition and suddenly rises back to the radiation value $c_s=1/\sqrt{3}$
after the transition. A significant decrease in the effective
sound speed $c_s$ during the cosmological QCD transition was pointed out 
by Jedamzik \cite{Jedamzik} independently.
Pressure varies continuously and goes below the ideal radiation fluid value
$p=\rho/3$, but stays positive.

As the sound speed drops to zero, the
restoring force for acoustic oscillations vanishes and density perturbations
for subhorizon modes fall freely. The fluid velocity stays constant during
this free fall. Perturbations of shorter wavelengths have higher velocities
at the beginning of the transition, and thus grow proportional to wavenumber
$k$ during the phase transition. The primordial Harrison-Zel'dovich
spectrum of density perturbations is amplified on subhorizon scales.
It develops peaks which grow, at most, linearly in wavenumber, see Fig.~2.
We used a fit to the QCD equation of state obtained on the lattice 
\cite{Boyd,MILC97}. The spectrum of density perturbations
on superhorizon scales, $\lambda > R_{\rm H}$, is unaffected.
At $T\sim 1$ MeV the neutrinos decouple from the radiation fluid.
During this decoupling the large peaks in the radiation spectrum
are wiped out by collisional damping.

Today the universe is dominated by dark matter, most likely cold dark 
matter (CDM). If CDM is kinetically decoupled from the radiation fluid
at the QCD transition, the density perturbations in CDM do not suffer
from the neutrino damping. This is the case for primordial black 
holes or axions, but not for supersymmetric dark matter.
At the time of the QCD transition the energy density of CDM is small, i.e.\ 
$\rho^{\rm CDM}(T_\star) \sim 10^{-8} \rho^{\rm RAD}(T_\star)$.
CDM falls into the potential wells provided by the dominant
radiation fluid. Thus, the CDM spectrum is amplified on subhorizon
scales, see Fig.~2. The peaks in the CDM spectrum go nonlinear
shortly after radiation-matter equality. This leads to the 
formation of CDM clumps with
mass $< 10^{-10} M_\odot$. Especially the clumping of axions has important
implications for axion searches with strong magnetic fields \cite{Sikivie}.
If the QCD transition is strong enough, there is a chance that these
clumps could be detected by gravitational femtolensing \cite{femtolensing}.

The vanishing of the sound speed during the coexistence phase also leads to
interesting gravitational effects, as pointed out by Jedamzik
\cite{Jedamzik}. He found that the probability to form black holes
is enhanced during the QCD transition due to the vanishing pressure gradients.
These black holes have masses $\sim 1 M_{\odot}$. He claims that
black holes from the QCD transition could account for the massive compact 
halo objects (MACHO's)
observed by microlensing \cite{microlensing} in the halo of our galaxy. 
However, for standard models of
structure formation (standard CDM, $\Lambda$CDM, etc.) the
amplitude of fluctuations at the QCD horizon crossing scale is not big
enough to produce a cosmologically relevant amount of black holes.
A fine tuned primordial spectrum would be necessary.

\section{Neutron stars --- quark stars}

In this section I concentrate on observable aspects of compact stars.
A neutron star mostly consists of neutrons, but due to weak 
interactions it has an admixture of protons and electrons 
($\mu_{\rm n} = \mu_{\rm p} + \mu_{\rm e}$). The equation of state is a 
function of two chemical potentials (if we neglect temperature), because 
there are two conserved charges, baryon number and electric charge. If the 
chemical potentials are high enough, heavier particles contribute to the star's 
composition ($\mu$, $\pi$, $\Lambda$, etc.). Besides the hadron core (with 
energy densities $\rho \sim 10^{15}$ g/cm$^3 \sim \Lambda_{\rm QCD}^4$) 
the star has a crust of iron (with $\rho < 10^{13}$ g/cm$^3$). An extensive 
discussion of the structure and properties of compact stars can be found in 
\cite{GlendenningBook}.

Asymptotic freedom of QCD suggests that at high baryon chemical potential 
a deconfinement transition happens. The core of the neutron star may consist 
of a phase of free quarks and gluons \cite{Itho}. An extended mixed phase 
might exist, because charge has to be conserved globally, but not locally 
\cite{Glendenning92}. Such objects are called hybrid stars. If strange matter
(made of u, d, and s quarks) is absolutely stable the star is called quark 
star (or strange star). The details of the hybrid and quark stars'
composition and structure strongly depend on the equations of state of QCD and 
nuclear matter \cite{Glendenning,quarkstar}. 

An important feature of neutron stars is their strong magnetic field
(typically $10^7$ T). When a neutron star rotates fast enough, 
the magnetic dipole radiates. The neutron star can be observed
as pulsar (more than $730$ have been detected). Some pulsars stand 
alone and some come in binaries. For the binary systems the mass of the pulsar
can be obtained (about $20$ masses are known). The mean mass is $\sim 1.4 
M_\odot$\cite{Thorsett}, the maximum mass is $\sim 1.8 M_\odot$ (Vela X-1). 
The pulsar ages are known if the supernova from which the pulsar formed has 
been observed (Crab) and/or some remnants from the supernova explosion
are observable. Another way to estimate the pulsar age starts from the 
assumption that the spin-down is due to the magnetic dipole only. In this case
$\tau_{\rm dipole} \equiv \dot{P}/(2P)$, where $P$ is the period of the pulsar.
Surface temperatures of pulsars have been measured
for a couple of pulsars by the X-ray satellites ROSAT and ASCA
\cite{Greiveldinger}. Fig.~3 shows the surface temperature and dipole
age for four pulsars.

\begin{figure}
\begin{center}
\epsfig{figure=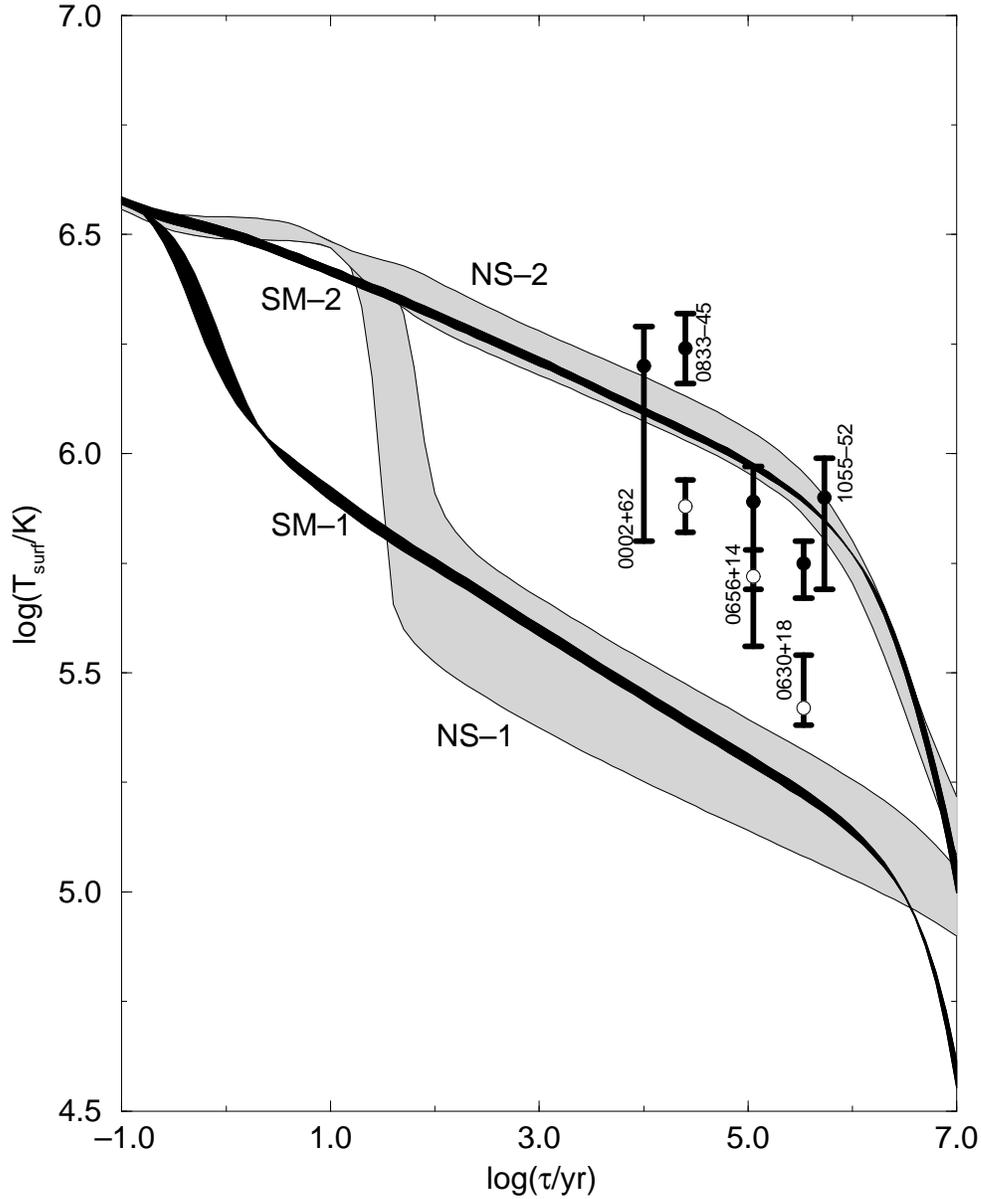,width=0.8\linewidth}
\end{center}
\vspace{-1 cm}
\caption{
The cooling behavior of non-superfluid strange matter stars SM-1 (lower
solid band) and SM-2 (upper solid band) and non-superfluid neutron
stars NS-1 (lower shaded band) and NS-2 (upper shaded band). The
surface temperatures of several observed neutron stars obtained for a
blackbody (magnetic hydrogen) atmosphere are marked with error bars
with solid (hollow)  circles. Courtesy of F. Weber \cite{Weber}.
}
\vspace{- \baselineskip}
\end{figure}
Can observations decide which is the correct model for neutron stars? 
Of course, every model should include masses as large as the heaviest 
observed pulsar. Another prediction from the models is the cooling curve. 
After a supernova explosion the remaining central star is hot and has 
to cool by emitting neutrinos mainly. The cooling rate crucially depends 
on the kind of allowed weak (so-called Urca) processes, 
like $p + e \to n + \nu_{\rm e}$ (direct Urca). If the fraction of 
protons is small, the triangle inequality between 
the Fermi momenta, $p_{\rm F}^{\rm e} + p_{\rm F}^{\rm p} \geq 
p_{\rm F}^{\rm n}$, is violated and the direct Urca processes are 
forbidden. In this case only modified Urca processes and/or
neutrino bremsstrahlung can occur and cooling is slower. 
For neutron stars with a core of nucleons
below some critical proton fraction, the slow cooling scenario applies. 
Higher densities or more exotic matter (hyperons, kaon condensate, etc.) 
enhance the cooling \cite{Boguta,Pethick}. The same happens
for a quark core \cite{Iwamoto}. 
A recent calculation for competing models [strange star (SM) and 
neutron star (NS), both with enhanced (1) and standard cooling (2)] 
has been performed by Schaab et al. \cite{Schaab}, see Fig.~3. 
A comparison with observations seems to slightly prefer the suppression of 
direct Urca processes (models SM-2 and NS-2 fit better than
model SM-1 and NS-1 in Fig.~3). Another set of calculations 
includes superfluid phases in neutron stars and strange stars. These models 
do not show significant differences in their cooling behavior. 

These findings might support the ideas of Alford, Rajagopal, and 
Wilczek \cite{Wilczek}, who argue that a color-flavor locked state
might exist in the core of a quark star. Due to the locking of flavor, 
the direct Urca processes should be suppressed (similar to the superfluid 
models of \cite{Schaab}). The color-flavor locked  phase
exists, if there is some attractive force between two quarks 
(there is an attractive one-gluon-exchange diagram, but 
the next order is repulsive; a non-perturbative answer is 
needed). Other observational consequences of the color-flavor locked
state remain to be investigated \cite{Wtalk}.

\begin{figure}
\begin{center}
\epsfig{figure=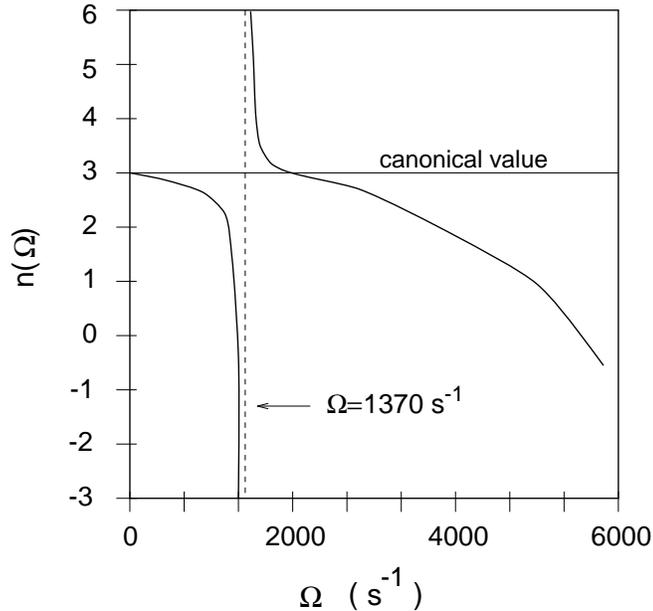,width=0.5\linewidth,angle=-90}
\end{center}
\vspace{-1 cm}
\caption{
The braking index $n$ as a function of rotational frequency for a hybrid
star. The dip at frequencies $\Omega \sim 1370$ s$^{-1}$ originates from
quark deconfinement. The overall reduction of $n$ below 3 is due to
the frequency dependence of the moment of inertia and, 
thus, is independent of whether or not quark deconfinement takes place 
in pulsars. Courtesy of F. Weber \cite{Weber}.
}
\vspace{- \baselineskip}
\end{figure}
Pulsar timing is one of the most precise experiments in nature.
Not only the angular velocity of a pulsar can be measured, but also its 
first and second derivatives. This allows to look for phase transitions
inside the pulsar. This important effect has been discovered recently by
Glendenning, Pei, and Weber \cite{Glendenning}. Pulsars are 
decelerated by electromagnetic emission and by a wind of electrons and 
positrons. As a pulsar slows down, its central density increases.
At some critical density a phase transition might occur. The radius of 
the star shrinks, and the moment of inertia decreases. Since angular momentum 
is conserved, this causes a period of spin-up, which lasts
about $2\times 10^7$ years (a $1/50$th of the pulsar's spin-down time). The most
interesting observable is the braking index $n \equiv \ddot{\Omega}\Omega/
\dot{\Omega}^2$ ($\Omega$ is the angular velocity), which grows to 
large values $|n| \gg 3$ (see Fig.~4). This would be a unique signature 
of a phase transition. The prospects to actually measure such spin-ups
are reasonably good. About $1/50$th of all pulsars might pass through such a 
phase currently. So far, the braking index has been measured for a few 
pulsars only. 

Let me finally comment on the ultimate heavy 'ion' collision, a collision of
two neutron stars. It has been suggested that collisions of neutron stars
might be the source of short gamma-ray bursts \cite{Katz}. A large number of 
gamma-ray bursts (GRB), the most energetic events in the universe,
have been detected in the recent years by the 
BATSE instrument on the CGRO spacecraft \cite{BATSE}. Their distribution is
isotropic across the sky and they are placed at cosmological distances,
as demonstrated by the recent discoveries of X-ray, optical, and 
radio transients \cite{Beppo}. For the GRB of 14 December 1997 the host galaxy
was identified at a redshift of $z=3.42$ \cite{highz}. This large distance 
implies that about $3 \times 10^{53}$ erg, corresponding to a rest-mass energy
of $0.2 M_\odot$, have been released in $\gamma$-rays. Energy
released in neutrinos or gravitational waves is not included in this number.
The observed durations of GRBs vary from several ms to $10^3$ s. The observed
photons have energies up to $26$ GeV \cite{highe}. 

To understand the origin of GRBs several mechanisms have been proposed.
The models have in common that the observed GRBs stem from a
relativistic fireball (with Lorentz factor $> 100$) \cite{fireball}.
The so-called central engine of the fireball is unknown. A promising
scenario is the merging of two neutron stars \cite{Eichler}. All neutron
star binaries lose energy by gravitational radiation, finally they
spiral in and merge. Such events
are expected to happen at a rate $10^{-6}$ per year per galaxy, which allows
to observe about one event per day. 

A even more dramatic scenario has been suggested in Ref.~\cite{Katz}. 
The collision of two neutron stars in very dense clusters of stars
could account for short (ms) GRBs. Numerical simulations of neutron star 
collisions have been performed in \cite{Janka}. The authors find an extremely 
luminous outburst of neutrinos. From the annihilation of neutrinos 
and anti-neutrinos a fireball of electrons and positrons with the right
amount of energy is formed. However, they find a large baryon contamination 
of the fireball, too large to explain the observed GRB. However, the
dependence of this result on the equation of state has not been investigated.
A soft point in the equation of state certainly would change the baryon 
contamination. As in the cosmological QCD transition a first order 
transition could give rise to small sound speed, and pressure gradients 
could be much smaller than in the simulations \cite{Janka}, giving rise to 
slower expansion of the baryon fluid.

\section{Conclusions}

The success of standard big bang nucleosynthesis suggests that
the cosmological QCD transition is either a smooth crossover, 
or, in the case of a first order transition, latent heat $l$ and surface 
tension $\sigma$ are restricted by $v^2 \sigma^3/(l^2 T_\star) < 100$. 
This bound follows from $d_{\rm nucl} < d_{\rm proton\ diffusion} \sim 
10^{-4} R_{\rm H}$, using the analytical results for $d_{\rm nucl}$ of 
\cite{Ignatius}. $v$ is the velocity at which the released latent heat is 
transported into the quark phase, a good guess might be $v \sim 0.1 c$.
This bound is consistent with lattice QCD results \cite{Iwasaki2}. 

For a cosmological QCD transition of first order we expect clumps in CDM
that is kinetically decoupled at the QCD scale (axions or primordial 
black holes) \cite{SSW}. This would reduce the chances to detect axions 
in ongoing searches \cite{Sikivie}. A further study of the evolution of 
clumped CDM is needed to give quantitative answers, a better knowledge 
of the QCD equation of state (from lattice QCD) is needed to predict the 
size of these clumps.

A QCD transition inside a neutron star could be detected by the severe 
spin-up of pulsars \cite{Glendenning}. Pulsar timing needs
long periods of observation, thus, patience is required. 

Cooling curves of pulsars might be a tool to rule out some of the equations
of state for matter at high baryon density. Progress in the calculation of
equations of state at high baryon chemical potential on the lattice
is of major importance for a better understanding of the most compact
stars in the universe.

The collision of two neutron stars might give new insight into the equation of
state in neutron stars. These collisions might have been observed in 
short gamma-ray bursts \cite{Katz}. 

\section*{Acknowledgment}

I am grateful to M. Hofmann, M. Hanauske, J. Ignatius, F. Weber,
and P. Widerin for discussions and/or comments on the manuscript.
I thank the Alexander von Humboldt foundation for financial support.

\end{document}

%% file: om.tex
\setlength{\unitlength}{0.240900pt}
\ifx\plotpoint\undefined\newsavebox{\plotpoint}\fi
\sbox{\plotpoint}{\rule[-0.200pt]{0.400pt}{0.400pt}}%
\begin{picture}(1500,900)(0,0)
\font\gnuplot=cmr10 at 10pt
\gnuplot
\sbox{\plotpoint}{\rule[-0.200pt]{0.400pt}{0.400pt}}%
\put(176.0,113.0){\rule[-0.200pt]{4.818pt}{0.400pt}}
\put(154,113){\makebox(0,0)[r]{$0.6$}}
\put(1416.0,113.0){\rule[-0.200pt]{4.818pt}{0.400pt}}
\put(176.0,266.0){\rule[-0.200pt]{4.818pt}{0.400pt}}
\put(154,266){\makebox(0,0)[r]{$0.7$}}
\put(1416.0,266.0){\rule[-0.200pt]{4.818pt}{0.400pt}}
\put(176.0,419.0){\rule[-0.200pt]{4.818pt}{0.400pt}}
\put(154,419){\makebox(0,0)[r]{$0.8$}}
\put(1416.0,419.0){\rule[-0.200pt]{4.818pt}{0.400pt}}
\put(176.0,571.0){\rule[-0.200pt]{4.818pt}{0.400pt}}
\put(154,571){\makebox(0,0)[r]{$0.9$}}
\put(1416.0,571.0){\rule[-0.200pt]{4.818pt}{0.400pt}}
\put(176.0,724.0){\rule[-0.200pt]{4.818pt}{0.400pt}}
\put(154,724){\makebox(0,0)[r]{$1$}}
\put(1416.0,724.0){\rule[-0.200pt]{4.818pt}{0.400pt}}
\put(176.0,877.0){\rule[-0.200pt]{4.818pt}{0.400pt}}
\put(154,877){\makebox(0,0)[r]{$1.1$}}
\put(1416.0,877.0){\rule[-0.200pt]{4.818pt}{0.400pt}}
\put(176.0,113.0){\rule[-0.200pt]{0.400pt}{2.409pt}}
\put(176.0,867.0){\rule[-0.200pt]{0.400pt}{2.409pt}}
\put(231.0,113.0){\rule[-0.200pt]{0.400pt}{2.409pt}}
\put(231.0,867.0){\rule[-0.200pt]{0.400pt}{2.409pt}}
\put(271.0,113.0){\rule[-0.200pt]{0.400pt}{2.409pt}}
\put(271.0,867.0){\rule[-0.200pt]{0.400pt}{2.409pt}}
\put(301.0,113.0){\rule[-0.200pt]{0.400pt}{2.409pt}}
\put(301.0,867.0){\rule[-0.200pt]{0.400pt}{2.409pt}}
\put(326.0,113.0){\rule[-0.200pt]{0.400pt}{2.409pt}}
\put(326.0,867.0){\rule[-0.200pt]{0.400pt}{2.409pt}}
\put(347.0,113.0){\rule[-0.200pt]{0.400pt}{2.409pt}}
\put(347.0,867.0){\rule[-0.200pt]{0.400pt}{2.409pt}}
\put(366.0,113.0){\rule[-0.200pt]{0.400pt}{2.409pt}}
\put(366.0,867.0){\rule[-0.200pt]{0.400pt}{2.409pt}}
\put(382.0,113.0){\rule[-0.200pt]{0.400pt}{2.409pt}}
\put(382.0,867.0){\rule[-0.200pt]{0.400pt}{2.409pt}}
\put(396.0,113.0){\rule[-0.200pt]{0.400pt}{4.818pt}}
\put(396,68){\makebox(0,0){$10^{-8}$}}
\put(396.0,857.0){\rule[-0.200pt]{0.400pt}{4.818pt}}
\put(491.0,113.0){\rule[-0.200pt]{0.400pt}{2.409pt}}
\put(491.0,867.0){\rule[-0.200pt]{0.400pt}{2.409pt}}
\put(546.0,113.0){\rule[-0.200pt]{0.400pt}{2.409pt}}
\put(546.0,867.0){\rule[-0.200pt]{0.400pt}{2.409pt}}
\put(586.0,113.0){\rule[-0.200pt]{0.400pt}{2.409pt}}
\put(586.0,867.0){\rule[-0.200pt]{0.400pt}{2.409pt}}
\put(616.0,113.0){\rule[-0.200pt]{0.400pt}{2.409pt}}
\put(616.0,867.0){\rule[-0.200pt]{0.400pt}{2.409pt}}
\put(641.0,113.0){\rule[-0.200pt]{0.400pt}{2.409pt}}
\put(641.0,867.0){\rule[-0.200pt]{0.400pt}{2.409pt}}
\put(662.0,113.0){\rule[-0.200pt]{0.400pt}{2.409pt}}
\put(662.0,867.0){\rule[-0.200pt]{0.400pt}{2.409pt}}
\put(681.0,113.0){\rule[-0.200pt]{0.400pt}{2.409pt}}
\put(681.0,867.0){\rule[-0.200pt]{0.400pt}{2.409pt}}
\put(697.0,113.0){\rule[-0.200pt]{0.400pt}{2.409pt}}
\put(697.0,867.0){\rule[-0.200pt]{0.400pt}{2.409pt}}
\put(711.0,113.0){\rule[-0.200pt]{0.400pt}{4.818pt}}
\put(711,68){\makebox(0,0){$10^{-7}$}}
\put(711.0,857.0){\rule[-0.200pt]{0.400pt}{4.818pt}}
\put(806.0,113.0){\rule[-0.200pt]{0.400pt}{2.409pt}}
\put(806.0,867.0){\rule[-0.200pt]{0.400pt}{2.409pt}}
\put(861.0,113.0){\rule[-0.200pt]{0.400pt}{2.409pt}}
\put(861.0,867.0){\rule[-0.200pt]{0.400pt}{2.409pt}}
\put(901.0,113.0){\rule[-0.200pt]{0.400pt}{2.409pt}}
\put(901.0,867.0){\rule[-0.200pt]{0.400pt}{2.409pt}}
\put(931.0,113.0){\rule[-0.200pt]{0.400pt}{2.409pt}}
\put(931.0,867.0){\rule[-0.200pt]{0.400pt}{2.409pt}}
\put(956.0,113.0){\rule[-0.200pt]{0.400pt}{2.409pt}}
\put(956.0,867.0){\rule[-0.200pt]{0.400pt}{2.409pt}}
\put(977.0,113.0){\rule[-0.200pt]{0.400pt}{2.409pt}}
\put(977.0,867.0){\rule[-0.200pt]{0.400pt}{2.409pt}}
\put(996.0,113.0){\rule[-0.200pt]{0.400pt}{2.409pt}}
\put(996.0,867.0){\rule[-0.200pt]{0.400pt}{2.409pt}}
\put(1012.0,113.0){\rule[-0.200pt]{0.400pt}{2.409pt}}
\put(1012.0,867.0){\rule[-0.200pt]{0.400pt}{2.409pt}}
\put(1026.0,113.0){\rule[-0.200pt]{0.400pt}{4.818pt}}
\put(1026,68){\makebox(0,0){$10^{-6}$}}
\put(1026.0,857.0){\rule[-0.200pt]{0.400pt}{4.818pt}}
\put(1121.0,113.0){\rule[-0.200pt]{0.400pt}{2.409pt}}
\put(1121.0,867.0){\rule[-0.200pt]{0.400pt}{2.409pt}}
\put(1176.0,113.0){\rule[-0.200pt]{0.400pt}{2.409pt}}
\put(1176.0,867.0){\rule[-0.200pt]{0.400pt}{2.409pt}}
\put(1216.0,113.0){\rule[-0.200pt]{0.400pt}{2.409pt}}
\put(1216.0,867.0){\rule[-0.200pt]{0.400pt}{2.409pt}}
\put(1246.0,113.0){\rule[-0.200pt]{0.400pt}{2.409pt}}
\put(1246.0,867.0){\rule[-0.200pt]{0.400pt}{2.409pt}}
\put(1271.0,113.0){\rule[-0.200pt]{0.400pt}{2.409pt}}
\put(1271.0,867.0){\rule[-0.200pt]{0.400pt}{2.409pt}}
\put(1292.0,113.0){\rule[-0.200pt]{0.400pt}{2.409pt}}
\put(1292.0,867.0){\rule[-0.200pt]{0.400pt}{2.409pt}}
\put(1311.0,113.0){\rule[-0.200pt]{0.400pt}{2.409pt}}
\put(1311.0,867.0){\rule[-0.200pt]{0.400pt}{2.409pt}}
\put(1327.0,113.0){\rule[-0.200pt]{0.400pt}{2.409pt}}
\put(1327.0,867.0){\rule[-0.200pt]{0.400pt}{2.409pt}}
\put(1341.0,113.0){\rule[-0.200pt]{0.400pt}{4.818pt}}
\put(1341,68){\makebox(0,0){$10^{-5}$}}
\put(1341.0,857.0){\rule[-0.200pt]{0.400pt}{4.818pt}}
\put(1436.0,113.0){\rule[-0.200pt]{0.400pt}{2.409pt}}
\put(1436.0,867.0){\rule[-0.200pt]{0.400pt}{2.409pt}}
\put(176.0,113.0){\rule[-0.200pt]{303.534pt}{0.400pt}}
\put(1436.0,113.0){\rule[-0.200pt]{0.400pt}{184.048pt}}
\put(176.0,877.0){\rule[-0.200pt]{303.534pt}{0.400pt}}
\put(806,23){\makebox(0,0){\large Hz}}
\put(429,225){\makebox(0,0){pulsar timing}}
\put(736,225){\makebox(0,0)[l]{\large $f_\star$}}
\put(711,790){\makebox(0,0)[l]
{\LARGE $\Omega_{\rm gw}(f)\over\Omega_{\rm gw}(f\ll f_\star)$}}
\put(176.0,113.0){\rule[-0.200pt]{0.400pt}{184.048pt}}
\put(747,189){\vector(0,-1){45}}
\put(396,189){\vector(-1,0){125}}
\put(396,189){\vector(1,0){190}}
\put(1306,812){\makebox(0,0)[r]{bag model}}
\put(1328.0,812.0){\rule[-0.200pt]{15.899pt}{0.400pt}}
\put(360,722.67){\rule{0.964pt}{0.400pt}}
\multiput(360.00,723.17)(2.000,-1.000){2}{\rule{0.482pt}{0.400pt}}
\put(176.0,724.0){\rule[-0.200pt]{44.326pt}{0.400pt}}
\put(421,721.67){\rule{1.204pt}{0.400pt}}
\multiput(421.00,722.17)(2.500,-1.000){2}{\rule{0.602pt}{0.400pt}}
\put(364.0,723.0){\rule[-0.200pt]{13.731pt}{0.400pt}}
\put(452,720.67){\rule{1.204pt}{0.400pt}}
\multiput(452.00,721.17)(2.500,-1.000){2}{\rule{0.602pt}{0.400pt}}
\put(426.0,722.0){\rule[-0.200pt]{6.263pt}{0.400pt}}
\put(474,719.67){\rule{1.204pt}{0.400pt}}
\multiput(474.00,720.17)(2.500,-1.000){2}{\rule{0.602pt}{0.400pt}}
\put(457.0,721.0){\rule[-0.200pt]{4.095pt}{0.400pt}}
\put(492,718.67){\rule{0.964pt}{0.400pt}}
\multiput(492.00,719.17)(2.000,-1.000){2}{\rule{0.482pt}{0.400pt}}
\put(479.0,720.0){\rule[-0.200pt]{3.132pt}{0.400pt}}
\put(505,717.67){\rule{1.204pt}{0.400pt}}
\multiput(505.00,718.17)(2.500,-1.000){2}{\rule{0.602pt}{0.400pt}}
\put(496.0,719.0){\rule[-0.200pt]{2.168pt}{0.400pt}}
\put(514,716.67){\rule{0.964pt}{0.400pt}}
\multiput(514.00,717.17)(2.000,-1.000){2}{\rule{0.482pt}{0.400pt}}
\put(510.0,718.0){\rule[-0.200pt]{0.964pt}{0.400pt}}
\put(527,715.67){\rule{1.204pt}{0.400pt}}
\multiput(527.00,716.17)(2.500,-1.000){2}{\rule{0.602pt}{0.400pt}}
\put(532,714.67){\rule{0.964pt}{0.400pt}}
\multiput(532.00,715.17)(2.000,-1.000){2}{\rule{0.482pt}{0.400pt}}
\put(518.0,717.0){\rule[-0.200pt]{2.168pt}{0.400pt}}
\put(540,713.67){\rule{1.204pt}{0.400pt}}
\multiput(540.00,714.17)(2.500,-1.000){2}{\rule{0.602pt}{0.400pt}}
\put(536.0,715.0){\rule[-0.200pt]{0.964pt}{0.400pt}}
\put(549,712.67){\rule{1.204pt}{0.400pt}}
\multiput(549.00,713.17)(2.500,-1.000){2}{\rule{0.602pt}{0.400pt}}
\put(554,711.67){\rule{0.964pt}{0.400pt}}
\multiput(554.00,712.17)(2.000,-1.000){2}{\rule{0.482pt}{0.400pt}}
\put(558,710.67){\rule{0.964pt}{0.400pt}}
\multiput(558.00,711.17)(2.000,-1.000){2}{\rule{0.482pt}{0.400pt}}
\put(545.0,714.0){\rule[-0.200pt]{0.964pt}{0.400pt}}
\put(567,709.67){\rule{0.964pt}{0.400pt}}
\multiput(567.00,710.17)(2.000,-1.000){2}{\rule{0.482pt}{0.400pt}}
\put(571,708.67){\rule{1.204pt}{0.400pt}}
\multiput(571.00,709.17)(2.500,-1.000){2}{\rule{0.602pt}{0.400pt}}
\put(576,707.67){\rule{0.964pt}{0.400pt}}
\multiput(576.00,708.17)(2.000,-1.000){2}{\rule{0.482pt}{0.400pt}}
\put(580,706.67){\rule{1.204pt}{0.400pt}}
\multiput(580.00,707.17)(2.500,-1.000){2}{\rule{0.602pt}{0.400pt}}
\put(585,705.17){\rule{0.900pt}{0.400pt}}
\multiput(585.00,706.17)(2.132,-2.000){2}{\rule{0.450pt}{0.400pt}}
\put(589,703.67){\rule{0.964pt}{0.400pt}}
\multiput(589.00,704.17)(2.000,-1.000){2}{\rule{0.482pt}{0.400pt}}
\put(593,702.67){\rule{1.204pt}{0.400pt}}
\multiput(593.00,703.17)(2.500,-1.000){2}{\rule{0.602pt}{0.400pt}}
\put(598,701.67){\rule{0.964pt}{0.400pt}}
\multiput(598.00,702.17)(2.000,-1.000){2}{\rule{0.482pt}{0.400pt}}
\put(602,700.17){\rule{1.100pt}{0.400pt}}
\multiput(602.00,701.17)(2.717,-2.000){2}{\rule{0.550pt}{0.400pt}}
\put(607,698.17){\rule{0.900pt}{0.400pt}}
\multiput(607.00,699.17)(2.132,-2.000){2}{\rule{0.450pt}{0.400pt}}
\put(611,696.67){\rule{0.964pt}{0.400pt}}
\multiput(611.00,697.17)(2.000,-1.000){2}{\rule{0.482pt}{0.400pt}}
\put(615,695.17){\rule{1.100pt}{0.400pt}}
\multiput(615.00,696.17)(2.717,-2.000){2}{\rule{0.550pt}{0.400pt}}
\put(620,693.17){\rule{0.900pt}{0.400pt}}
\multiput(620.00,694.17)(2.132,-2.000){2}{\rule{0.450pt}{0.400pt}}
\put(624,691.17){\rule{1.100pt}{0.400pt}}
\multiput(624.00,692.17)(2.717,-2.000){2}{\rule{0.550pt}{0.400pt}}
\put(629,689.17){\rule{0.900pt}{0.400pt}}
\multiput(629.00,690.17)(2.132,-2.000){2}{\rule{0.450pt}{0.400pt}}
\put(633,687.17){\rule{0.900pt}{0.400pt}}
\multiput(633.00,688.17)(2.132,-2.000){2}{\rule{0.450pt}{0.400pt}}
\multiput(637.00,685.95)(0.909,-0.447){3}{\rule{0.767pt}{0.108pt}}
\multiput(637.00,686.17)(3.409,-3.000){2}{\rule{0.383pt}{0.400pt}}
\put(642,682.17){\rule{0.900pt}{0.400pt}}
\multiput(642.00,683.17)(2.132,-2.000){2}{\rule{0.450pt}{0.400pt}}
\multiput(646.00,680.95)(0.909,-0.447){3}{\rule{0.767pt}{0.108pt}}
\multiput(646.00,681.17)(3.409,-3.000){2}{\rule{0.383pt}{0.400pt}}
\multiput(651.00,677.95)(0.685,-0.447){3}{\rule{0.633pt}{0.108pt}}
\multiput(651.00,678.17)(2.685,-3.000){2}{\rule{0.317pt}{0.400pt}}
\multiput(655.00,674.95)(0.685,-0.447){3}{\rule{0.633pt}{0.108pt}}
\multiput(655.00,675.17)(2.685,-3.000){2}{\rule{0.317pt}{0.400pt}}
\multiput(659.00,671.95)(0.909,-0.447){3}{\rule{0.767pt}{0.108pt}}
\multiput(659.00,672.17)(3.409,-3.000){2}{\rule{0.383pt}{0.400pt}}
\multiput(664.00,668.94)(0.481,-0.468){5}{\rule{0.500pt}{0.113pt}}
\multiput(664.00,669.17)(2.962,-4.000){2}{\rule{0.250pt}{0.400pt}}
\multiput(668.00,664.95)(0.909,-0.447){3}{\rule{0.767pt}{0.108pt}}
\multiput(668.00,665.17)(3.409,-3.000){2}{\rule{0.383pt}{0.400pt}}
\multiput(673.00,661.94)(0.481,-0.468){5}{\rule{0.500pt}{0.113pt}}
\multiput(673.00,662.17)(2.962,-4.000){2}{\rule{0.250pt}{0.400pt}}
\multiput(677.00,657.94)(0.627,-0.468){5}{\rule{0.600pt}{0.113pt}}
\multiput(677.00,658.17)(3.755,-4.000){2}{\rule{0.300pt}{0.400pt}}
\multiput(682.60,652.51)(0.468,-0.627){5}{\rule{0.113pt}{0.600pt}}
\multiput(681.17,653.75)(4.000,-3.755){2}{\rule{0.400pt}{0.300pt}}
\multiput(686.00,648.94)(0.481,-0.468){5}{\rule{0.500pt}{0.113pt}}
\multiput(686.00,649.17)(2.962,-4.000){2}{\rule{0.250pt}{0.400pt}}
\multiput(690.00,644.93)(0.487,-0.477){7}{\rule{0.500pt}{0.115pt}}
\multiput(690.00,645.17)(3.962,-5.000){2}{\rule{0.250pt}{0.400pt}}
\multiput(695.60,638.51)(0.468,-0.627){5}{\rule{0.113pt}{0.600pt}}
\multiput(694.17,639.75)(4.000,-3.755){2}{\rule{0.400pt}{0.300pt}}
\multiput(699.59,633.59)(0.477,-0.599){7}{\rule{0.115pt}{0.580pt}}
\multiput(698.17,634.80)(5.000,-4.796){2}{\rule{0.400pt}{0.290pt}}
\multiput(704.60,627.09)(0.468,-0.774){5}{\rule{0.113pt}{0.700pt}}
\multiput(703.17,628.55)(4.000,-4.547){2}{\rule{0.400pt}{0.350pt}}
\multiput(708.60,621.09)(0.468,-0.774){5}{\rule{0.113pt}{0.700pt}}
\multiput(707.17,622.55)(4.000,-4.547){2}{\rule{0.400pt}{0.350pt}}
\multiput(712.59,615.59)(0.477,-0.599){7}{\rule{0.115pt}{0.580pt}}
\multiput(711.17,616.80)(5.000,-4.796){2}{\rule{0.400pt}{0.290pt}}
\multiput(717.60,608.68)(0.468,-0.920){5}{\rule{0.113pt}{0.800pt}}
\multiput(716.17,610.34)(4.000,-5.340){2}{\rule{0.400pt}{0.400pt}}
\multiput(721.59,602.26)(0.477,-0.710){7}{\rule{0.115pt}{0.660pt}}
\multiput(720.17,603.63)(5.000,-5.630){2}{\rule{0.400pt}{0.330pt}}
\multiput(726.60,594.68)(0.468,-0.920){5}{\rule{0.113pt}{0.800pt}}
\multiput(725.17,596.34)(4.000,-5.340){2}{\rule{0.400pt}{0.400pt}}
\multiput(730.60,587.26)(0.468,-1.066){5}{\rule{0.113pt}{0.900pt}}
\multiput(729.17,589.13)(4.000,-6.132){2}{\rule{0.400pt}{0.450pt}}
\multiput(734.59,579.93)(0.477,-0.821){7}{\rule{0.115pt}{0.740pt}}
\multiput(733.17,581.46)(5.000,-6.464){2}{\rule{0.400pt}{0.370pt}}
\multiput(739.60,570.85)(0.468,-1.212){5}{\rule{0.113pt}{1.000pt}}
\multiput(738.17,572.92)(4.000,-6.924){2}{\rule{0.400pt}{0.500pt}}
\multiput(743.59,562.60)(0.477,-0.933){7}{\rule{0.115pt}{0.820pt}}
\multiput(742.17,564.30)(5.000,-7.298){2}{\rule{0.400pt}{0.410pt}}
\multiput(748.60,552.85)(0.468,-1.212){5}{\rule{0.113pt}{1.000pt}}
\multiput(747.17,554.92)(4.000,-6.924){2}{\rule{0.400pt}{0.500pt}}
\multiput(752.59,544.26)(0.477,-1.044){7}{\rule{0.115pt}{0.900pt}}
\multiput(751.17,546.13)(5.000,-8.132){2}{\rule{0.400pt}{0.450pt}}
\multiput(757.60,533.43)(0.468,-1.358){5}{\rule{0.113pt}{1.100pt}}
\multiput(756.17,535.72)(4.000,-7.717){2}{\rule{0.400pt}{0.550pt}}
\multiput(761.60,523.02)(0.468,-1.505){5}{\rule{0.113pt}{1.200pt}}
\multiput(760.17,525.51)(4.000,-8.509){2}{\rule{0.400pt}{0.600pt}}
\multiput(765.59,512.93)(0.477,-1.155){7}{\rule{0.115pt}{0.980pt}}
\multiput(764.17,514.97)(5.000,-8.966){2}{\rule{0.400pt}{0.490pt}}
\multiput(770.60,501.02)(0.468,-1.505){5}{\rule{0.113pt}{1.200pt}}
\multiput(769.17,503.51)(4.000,-8.509){2}{\rule{0.400pt}{0.600pt}}
\multiput(774.59,490.60)(0.477,-1.267){7}{\rule{0.115pt}{1.060pt}}
\multiput(773.17,492.80)(5.000,-9.800){2}{\rule{0.400pt}{0.530pt}}
\multiput(779.60,477.60)(0.468,-1.651){5}{\rule{0.113pt}{1.300pt}}
\multiput(778.17,480.30)(4.000,-9.302){2}{\rule{0.400pt}{0.650pt}}
\multiput(783.60,465.19)(0.468,-1.797){5}{\rule{0.113pt}{1.400pt}}
\multiput(782.17,468.09)(4.000,-10.094){2}{\rule{0.400pt}{0.700pt}}
\multiput(787.59,453.27)(0.477,-1.378){7}{\rule{0.115pt}{1.140pt}}
\multiput(786.17,455.63)(5.000,-10.634){2}{\rule{0.400pt}{0.570pt}}
\multiput(792.60,439.19)(0.468,-1.797){5}{\rule{0.113pt}{1.400pt}}
\multiput(791.17,442.09)(4.000,-10.094){2}{\rule{0.400pt}{0.700pt}}
\multiput(796.59,427.27)(0.477,-1.378){7}{\rule{0.115pt}{1.140pt}}
\multiput(795.17,429.63)(5.000,-10.634){2}{\rule{0.400pt}{0.570pt}}
\multiput(801.60,412.77)(0.468,-1.943){5}{\rule{0.113pt}{1.500pt}}
\multiput(800.17,415.89)(4.000,-10.887){2}{\rule{0.400pt}{0.750pt}}
\multiput(805.60,398.77)(0.468,-1.943){5}{\rule{0.113pt}{1.500pt}}
\multiput(804.17,401.89)(4.000,-10.887){2}{\rule{0.400pt}{0.750pt}}
\multiput(809.59,385.94)(0.477,-1.489){7}{\rule{0.115pt}{1.220pt}}
\multiput(808.17,388.47)(5.000,-11.468){2}{\rule{0.400pt}{0.610pt}}
\multiput(814.60,370.77)(0.468,-1.943){5}{\rule{0.113pt}{1.500pt}}
\multiput(813.17,373.89)(4.000,-10.887){2}{\rule{0.400pt}{0.750pt}}
\multiput(818.59,357.94)(0.477,-1.489){7}{\rule{0.115pt}{1.220pt}}
\multiput(817.17,360.47)(5.000,-11.468){2}{\rule{0.400pt}{0.610pt}}
\multiput(823.60,343.19)(0.468,-1.797){5}{\rule{0.113pt}{1.400pt}}
\multiput(822.17,346.09)(4.000,-10.094){2}{\rule{0.400pt}{0.700pt}}
\multiput(827.60,329.77)(0.468,-1.943){5}{\rule{0.113pt}{1.500pt}}
\multiput(826.17,332.89)(4.000,-10.887){2}{\rule{0.400pt}{0.750pt}}
\multiput(831.59,317.27)(0.477,-1.378){7}{\rule{0.115pt}{1.140pt}}
\multiput(830.17,319.63)(5.000,-10.634){2}{\rule{0.400pt}{0.570pt}}
\multiput(836.60,303.19)(0.468,-1.797){5}{\rule{0.113pt}{1.400pt}}
\multiput(835.17,306.09)(4.000,-10.094){2}{\rule{0.400pt}{0.700pt}}
\multiput(840.59,291.27)(0.477,-1.378){7}{\rule{0.115pt}{1.140pt}}
\multiput(839.17,293.63)(5.000,-10.634){2}{\rule{0.400pt}{0.570pt}}
\multiput(845.60,278.02)(0.468,-1.505){5}{\rule{0.113pt}{1.200pt}}
\multiput(844.17,280.51)(4.000,-8.509){2}{\rule{0.400pt}{0.600pt}}
\multiput(849.59,267.93)(0.477,-1.155){7}{\rule{0.115pt}{0.980pt}}
\multiput(848.17,269.97)(5.000,-8.966){2}{\rule{0.400pt}{0.490pt}}
\multiput(854.60,256.02)(0.468,-1.505){5}{\rule{0.113pt}{1.200pt}}
\multiput(853.17,258.51)(4.000,-8.509){2}{\rule{0.400pt}{0.600pt}}
\multiput(858.60,245.85)(0.468,-1.212){5}{\rule{0.113pt}{1.000pt}}
\multiput(857.17,247.92)(4.000,-6.924){2}{\rule{0.400pt}{0.500pt}}
\multiput(862.59,237.93)(0.477,-0.821){7}{\rule{0.115pt}{0.740pt}}
\multiput(861.17,239.46)(5.000,-6.464){2}{\rule{0.400pt}{0.370pt}}
\multiput(867.60,229.68)(0.468,-0.920){5}{\rule{0.113pt}{0.800pt}}
\multiput(866.17,231.34)(4.000,-5.340){2}{\rule{0.400pt}{0.400pt}}
\multiput(871.00,224.93)(0.487,-0.477){7}{\rule{0.500pt}{0.115pt}}
\multiput(871.00,225.17)(3.962,-5.000){2}{\rule{0.250pt}{0.400pt}}
\multiput(876.60,218.51)(0.468,-0.627){5}{\rule{0.113pt}{0.600pt}}
\multiput(875.17,219.75)(4.000,-3.755){2}{\rule{0.400pt}{0.300pt}}
\multiput(880.00,214.95)(0.685,-0.447){3}{\rule{0.633pt}{0.108pt}}
\multiput(880.00,215.17)(2.685,-3.000){2}{\rule{0.317pt}{0.400pt}}
\put(884,211.67){\rule{1.204pt}{0.400pt}}
\multiput(884.00,212.17)(2.500,-1.000){2}{\rule{0.602pt}{0.400pt}}
\put(562.0,711.0){\rule[-0.200pt]{1.204pt}{0.400pt}}
\put(893,211.67){\rule{1.204pt}{0.400pt}}
\multiput(893.00,211.17)(2.500,1.000){2}{\rule{0.602pt}{0.400pt}}
\put(898,213.17){\rule{0.900pt}{0.400pt}}
\multiput(898.00,212.17)(2.132,2.000){2}{\rule{0.450pt}{0.400pt}}
\multiput(902.00,215.60)(0.481,0.468){5}{\rule{0.500pt}{0.113pt}}
\multiput(902.00,214.17)(2.962,4.000){2}{\rule{0.250pt}{0.400pt}}
\multiput(906.00,219.59)(0.487,0.477){7}{\rule{0.500pt}{0.115pt}}
\multiput(906.00,218.17)(3.962,5.000){2}{\rule{0.250pt}{0.400pt}}
\multiput(911.60,224.00)(0.468,0.627){5}{\rule{0.113pt}{0.600pt}}
\multiput(910.17,224.00)(4.000,3.755){2}{\rule{0.400pt}{0.300pt}}
\multiput(915.59,229.00)(0.477,0.599){7}{\rule{0.115pt}{0.580pt}}
\multiput(914.17,229.00)(5.000,4.796){2}{\rule{0.400pt}{0.290pt}}
\multiput(920.60,235.00)(0.468,0.920){5}{\rule{0.113pt}{0.800pt}}
\multiput(919.17,235.00)(4.000,5.340){2}{\rule{0.400pt}{0.400pt}}
\multiput(924.59,242.00)(0.477,0.599){7}{\rule{0.115pt}{0.580pt}}
\multiput(923.17,242.00)(5.000,4.796){2}{\rule{0.400pt}{0.290pt}}
\multiput(929.60,248.00)(0.468,0.774){5}{\rule{0.113pt}{0.700pt}}
\multiput(928.17,248.00)(4.000,4.547){2}{\rule{0.400pt}{0.350pt}}
\multiput(933.60,254.00)(0.468,0.774){5}{\rule{0.113pt}{0.700pt}}
\multiput(932.17,254.00)(4.000,4.547){2}{\rule{0.400pt}{0.350pt}}
\multiput(937.00,260.59)(0.487,0.477){7}{\rule{0.500pt}{0.115pt}}
\multiput(937.00,259.17)(3.962,5.000){2}{\rule{0.250pt}{0.400pt}}
\multiput(942.00,265.60)(0.481,0.468){5}{\rule{0.500pt}{0.113pt}}
\multiput(942.00,264.17)(2.962,4.000){2}{\rule{0.250pt}{0.400pt}}
\put(946,269.17){\rule{1.100pt}{0.400pt}}
\multiput(946.00,268.17)(2.717,2.000){2}{\rule{0.550pt}{0.400pt}}
\put(951,271.17){\rule{0.900pt}{0.400pt}}
\multiput(951.00,270.17)(2.132,2.000){2}{\rule{0.450pt}{0.400pt}}
\put(889.0,212.0){\rule[-0.200pt]{0.964pt}{0.400pt}}
\put(959,271.67){\rule{1.204pt}{0.400pt}}
\multiput(959.00,272.17)(2.500,-1.000){2}{\rule{0.602pt}{0.400pt}}
\put(964,270.17){\rule{0.900pt}{0.400pt}}
\multiput(964.00,271.17)(2.132,-2.000){2}{\rule{0.450pt}{0.400pt}}
\multiput(968.00,268.95)(0.909,-0.447){3}{\rule{0.767pt}{0.108pt}}
\multiput(968.00,269.17)(3.409,-3.000){2}{\rule{0.383pt}{0.400pt}}
\multiput(973.00,265.95)(0.685,-0.447){3}{\rule{0.633pt}{0.108pt}}
\multiput(973.00,266.17)(2.685,-3.000){2}{\rule{0.317pt}{0.400pt}}
\multiput(977.00,262.95)(0.685,-0.447){3}{\rule{0.633pt}{0.108pt}}
\multiput(977.00,263.17)(2.685,-3.000){2}{\rule{0.317pt}{0.400pt}}
\multiput(981.00,259.95)(0.909,-0.447){3}{\rule{0.767pt}{0.108pt}}
\multiput(981.00,260.17)(3.409,-3.000){2}{\rule{0.383pt}{0.400pt}}
\put(986,256.17){\rule{0.900pt}{0.400pt}}
\multiput(986.00,257.17)(2.132,-2.000){2}{\rule{0.450pt}{0.400pt}}
\put(990,254.17){\rule{1.100pt}{0.400pt}}
\multiput(990.00,255.17)(2.717,-2.000){2}{\rule{0.550pt}{0.400pt}}
\put(955.0,273.0){\rule[-0.200pt]{0.964pt}{0.400pt}}
\put(1003,253.67){\rule{1.204pt}{0.400pt}}
\multiput(1003.00,253.17)(2.500,1.000){2}{\rule{0.602pt}{0.400pt}}
\put(1008,254.67){\rule{0.964pt}{0.400pt}}
\multiput(1008.00,254.17)(2.000,1.000){2}{\rule{0.482pt}{0.400pt}}
\put(1012,256.17){\rule{1.100pt}{0.400pt}}
\multiput(1012.00,255.17)(2.717,2.000){2}{\rule{0.550pt}{0.400pt}}
\put(1017,257.67){\rule{0.964pt}{0.400pt}}
\multiput(1017.00,257.17)(2.000,1.000){2}{\rule{0.482pt}{0.400pt}}
\put(1021,258.67){\rule{1.204pt}{0.400pt}}
\multiput(1021.00,258.17)(2.500,1.000){2}{\rule{0.602pt}{0.400pt}}
\put(1026,259.67){\rule{0.964pt}{0.400pt}}
\multiput(1026.00,259.17)(2.000,1.000){2}{\rule{0.482pt}{0.400pt}}
\put(995.0,254.0){\rule[-0.200pt]{1.927pt}{0.400pt}}
\put(1039,259.67){\rule{0.964pt}{0.400pt}}
\multiput(1039.00,260.17)(2.000,-1.000){2}{\rule{0.482pt}{0.400pt}}
\put(1030.0,261.0){\rule[-0.200pt]{2.168pt}{0.400pt}}
\put(1048,258.67){\rule{0.964pt}{0.400pt}}
\multiput(1048.00,259.17)(2.000,-1.000){2}{\rule{0.482pt}{0.400pt}}
\put(1043.0,260.0){\rule[-0.200pt]{1.204pt}{0.400pt}}
\put(1087,258.67){\rule{1.204pt}{0.400pt}}
\multiput(1087.00,258.17)(2.500,1.000){2}{\rule{0.602pt}{0.400pt}}
\put(1052.0,259.0){\rule[-0.200pt]{8.431pt}{0.400pt}}
\put(1100,258.67){\rule{1.204pt}{0.400pt}}
\multiput(1100.00,259.17)(2.500,-1.000){2}{\rule{0.602pt}{0.400pt}}
\put(1092.0,260.0){\rule[-0.200pt]{1.927pt}{0.400pt}}
\put(1105.0,259.0){\rule[-0.200pt]{79.738pt}{0.400pt}}
\sbox{\plotpoint}{\rule[-0.500pt]{1.000pt}{1.000pt}}%
\put(1306,767){\makebox(0,0)[r]{crossover}}
\multiput(1328,767)(20.756,0.000){4}{\usebox{\plotpoint}}
\put(1394,767){\usebox{\plotpoint}}
\put(176.00,724.00){\usebox{\plotpoint}}
\put(196.76,724.00){\usebox{\plotpoint}}
\multiput(199,724)(20.756,0.000){0}{\usebox{\plotpoint}}
\put(217.51,724.00){\usebox{\plotpoint}}
\multiput(224,724)(20.756,0.000){0}{\usebox{\plotpoint}}
\put(238.27,724.00){\usebox{\plotpoint}}
\put(259.02,724.00){\usebox{\plotpoint}}
\multiput(262,724)(20.756,0.000){0}{\usebox{\plotpoint}}
\put(279.78,724.00){\usebox{\plotpoint}}
\multiput(287,724)(20.756,0.000){0}{\usebox{\plotpoint}}
\put(300.53,723.96){\usebox{\plotpoint}}
\put(321.25,723.00){\usebox{\plotpoint}}
\multiput(325,723)(20.756,0.000){0}{\usebox{\plotpoint}}
\put(342.01,723.00){\usebox{\plotpoint}}
\put(362.76,723.00){\usebox{\plotpoint}}
\multiput(363,723)(20.694,-1.592){0}{\usebox{\plotpoint}}
\put(383.48,722.00){\usebox{\plotpoint}}
\multiput(388,722)(20.756,0.000){0}{\usebox{\plotpoint}}
\put(404.22,721.73){\usebox{\plotpoint}}
\put(424.91,720.08){\usebox{\plotpoint}}
\multiput(426,720)(20.756,0.000){0}{\usebox{\plotpoint}}
\put(445.64,719.45){\usebox{\plotpoint}}
\multiput(451,719)(20.694,-1.592){0}{\usebox{\plotpoint}}
\put(466.33,717.81){\usebox{\plotpoint}}
\put(486.93,715.32){\usebox{\plotpoint}}
\multiput(489,715)(20.514,-3.156){0}{\usebox{\plotpoint}}
\put(507.43,712.10){\usebox{\plotpoint}}
\multiput(514,711)(20.514,-3.156){0}{\usebox{\plotpoint}}
\put(527.91,708.77){\usebox{\plotpoint}}
\put(548.09,703.90){\usebox{\plotpoint}}
\multiput(552,703)(19.838,-6.104){0}{\usebox{\plotpoint}}
\put(567.90,697.79){\usebox{\plotpoint}}
\put(587.17,690.09){\usebox{\plotpoint}}
\multiput(590,689)(17.928,-10.458){0}{\usebox{\plotpoint}}
\put(605.37,680.18){\usebox{\plotpoint}}
\put(623.07,669.41){\usebox{\plotpoint}}
\put(639.34,656.55){\usebox{\plotpoint}}
\multiput(640,656)(15.844,-13.407){0}{\usebox{\plotpoint}}
\put(654.95,642.89){\usebox{\plotpoint}}
\put(668.89,627.52){\usebox{\plotpoint}}
\put(682.15,611.57){\usebox{\plotpoint}}
\put(694.31,594.76){\usebox{\plotpoint}}
\put(705.44,577.25){\usebox{\plotpoint}}
\multiput(716,561)(9.282,-18.564){2}{\usebox{\plotpoint}}
\put(735.41,522.75){\usebox{\plotpoint}}
\multiput(741,512)(9.004,-18.701){2}{\usebox{\plotpoint}}
\put(761.95,466.45){\usebox{\plotpoint}}
\multiput(766,457)(8.490,-18.940){2}{\usebox{\plotpoint}}
\put(786.50,409.24){\usebox{\plotpoint}}
\multiput(791,398)(8.490,-18.940){2}{\usebox{\plotpoint}}
\put(811.74,352.33){\usebox{\plotpoint}}
\put(820.58,333.55){\usebox{\plotpoint}}
\multiput(829,316)(10.559,-17.869){2}{\usebox{\plotpoint}}
\put(851.57,279.65){\usebox{\plotpoint}}
\put(865.14,264.01){\usebox{\plotpoint}}
\multiput(867,262)(17.065,-11.814){0}{\usebox{\plotpoint}}
\put(882.04,252.15){\usebox{\plotpoint}}
\put(901.84,246.49){\usebox{\plotpoint}}
\multiput(905,246)(20.684,1.724){0}{\usebox{\plotpoint}}
\put(922.38,248.24){\usebox{\plotpoint}}
\put(942.60,252.91){\usebox{\plotpoint}}
\multiput(943,253)(20.136,5.034){0}{\usebox{\plotpoint}}
\put(962.78,257.79){\usebox{\plotpoint}}
\multiput(968,259)(20.473,3.412){0}{\usebox{\plotpoint}}
\put(983.23,261.00){\usebox{\plotpoint}}
\put(1003.98,261.00){\usebox{\plotpoint}}
\multiput(1006,261)(20.756,0.000){0}{\usebox{\plotpoint}}
\put(1024.74,261.00){\usebox{\plotpoint}}
\multiput(1031,261)(20.684,-1.724){0}{\usebox{\plotpoint}}
\put(1045.45,260.00){\usebox{\plotpoint}}
\put(1066.18,260.78){\usebox{\plotpoint}}
\multiput(1069,261)(20.756,0.000){0}{\usebox{\plotpoint}}
\put(1086.93,261.00){\usebox{\plotpoint}}
\multiput(1094,261)(20.756,0.000){0}{\usebox{\plotpoint}}
\put(1107.68,261.00){\usebox{\plotpoint}}
\put(1128.44,261.00){\usebox{\plotpoint}}
\multiput(1132,261)(20.756,0.000){0}{\usebox{\plotpoint}}
\put(1149.19,261.00){\usebox{\plotpoint}}
\multiput(1157,261)(20.756,0.000){0}{\usebox{\plotpoint}}
\put(1169.95,261.00){\usebox{\plotpoint}}
\put(1190.70,261.00){\usebox{\plotpoint}}
\multiput(1195,261)(20.756,0.000){0}{\usebox{\plotpoint}}
\put(1211.46,261.00){\usebox{\plotpoint}}
\multiput(1220,261)(20.756,0.000){0}{\usebox{\plotpoint}}
\put(1232.21,261.00){\usebox{\plotpoint}}
\put(1252.97,261.00){\usebox{\plotpoint}}
\multiput(1258,261)(20.756,0.000){0}{\usebox{\plotpoint}}
\put(1273.73,261.00){\usebox{\plotpoint}}
\put(1294.48,261.00){\usebox{\plotpoint}}
\multiput(1295,261)(20.756,0.000){0}{\usebox{\plotpoint}}
\put(1315.24,261.00){\usebox{\plotpoint}}
\multiput(1321,261)(20.756,0.000){0}{\usebox{\plotpoint}}
\put(1333,261){\usebox{\plotpoint}}
\put(1350,261){\usebox{\plotpoint}}
\put(1370,261){\usebox{\plotpoint}}
\put(1390,261){\usebox{\plotpoint}}
\end{picture}

%% file: b.bbl
\begin{thebibliography}{99}
\bibitem{lattice}For recent reviews see K. Kanaya, Nucl. Phys. B
         (Proc. Suppl.) {\bf 47}, 144 (1996); E. Laermann, Nucl. Phys. B
         (Proc. Suppl.) {\bf 63}, 141 (1998).
\bibitem{Wilczek}M. Alford, K. Rajagopal, and F. Wilczek, Phys. Lett. B
         {\bf 422}, 247 (1998); e-print hep-ph/9804403 (1998); 
         R. Rapp et al., e-print hep-ph/9711396 (1997); J. Berges
         and K. Rajagopal, e-print hep-ph/9804233 (1998).
\bibitem{Wtalk}F. Wilczek, contributions to this volume.
\bibitem{Herrmann}N. Herrmann, contribution to this volume.
\bibitem{Witten}E. Witten, Phys. Rev. D {\bf 30}, 272 (1984).
\bibitem{IBBN}J. H. Applegate and C. J. Hogan, Phys. Rev. D {\bf 31}, 3037
         (1985); J. H. Applegate, C. J. Hogan, and R. J. Scherrer, Phys.
         Rev. D {\bf 35}, 1151 (1987); G. M. Fuller, G. J. Mathews, and C. R.
         Alcock, Phys. Rev. D {\bf 37}, 1380 (1988); R. A. Malaney and
         G. J. Mathews, Phys. Rep. {\bf 229}, 145 (1993).
\bibitem{Ignatius}J. Ignatius et al., Phys. Rev. D {\bf 49}, 3854 (1994);
         {\bf 50}, 3738 (1994).
\bibitem{Christiansen}M. B. Christiansen and J. Madsen, Phys. Rev. D {\bf 53},
         5446 (1996).
\bibitem{Iwasaki2}Y. Iwasaki et al., Phys. Rev. D {\bf 46}, 4657 (1992);
        {\bf 49}, 3540 (1994);
        B. Beinlich, F. Karsch, and A. Peikert, Phys. Lett. B {\bf 390}, 268
        (1997).
\bibitem{SSW}C. Schmid, D. J. Schwarz, and P. Widerin,
         Phys.\ Rev.\ Lett.\ {\bf 78}, 5468 (1997).
\bibitem{Jedamzik}K. Jedamzik, Phys. Rev. D {\bf 55}, 5871 (1997).
\bibitem{Schwarz}D. J. Schwarz, e-print gr-qc/9709027.
\bibitem{Barbour}I. Barbour, contribution to this volume.
\bibitem{Pethick}C. J. Pethick, Rev. Mod. Phys. {\bf 64}, 1133 (1992);
         C. Schaab et al., Nucl. Phys. {\bf A605}, 531 (1996).
\bibitem{Schaab}C. Schaab et al., Astrophys. J. {\bf 480}, L111 (1997).
\bibitem{Glendenning}N. K. Glendenning, S. Pei, and F. Weber, Phys. Rev. Lett.
         {\bf 79}, 1603 (1997).
\bibitem{Katz}J.I. Katz and L. M. Canel, Astrophys. J. {\bf 471}, 915 (1996);
         V. I. Dokuchaev and Yu. N. Eroshenko, Sov. Astron. Lett. {\bf 22},
         578 (1996).
\bibitem{BBN}For a recent review on big bang nucleosynthesis see 
         C. Caso et al., Eur. Phys. J. {\bf C3}, 1 (1998), 
         available on the PDG WWW pages (URL:{\tt http://pdg.lbl.gov/}). 
\bibitem{Weinberg}S. Weinberg, {\em Gravitation and Cosmology} (John Wiley
         \& Sons, New York, 1972).
\bibitem{MILC96}C. Bernard et al., Phys. Rev. D {\bf 54}, 4585 (1996).
\bibitem{Iwasaki}Y. Iwasaki et al., Z. Phys. C {\bf 71}, 343 (1996);
         Nucl. Phys. B (Proc. Suppl.) {\bf 47}, 515 (1996).
\bibitem{Brown}F. R. Brown et al., Phys. Rev. Lett. {\bf 20}, 2491 (1990).
\bibitem{Hackel}M. Hackel et al., Phys. Rev. D {\bf 46}, 5648 (1992).
\bibitem{Starobinskii}A. A. Starobinskii, Pis'ma Zh. Eksp. Teor. Fiz.
         {\bf 30}, 719 (1979) [JETP Lett. {\bf 30}, 682 (1979)].
\bibitem{pulsar}V. M. Kaspi, J. H. Taylor, and M. F. Ryba,
         Astrophys. J. {\bf 428}, 713 (1994);
         S. E. Thorsett and R. J. Dewey, Phys. Rev. D {\bf 53}, 3468 (1996);
         M. P. McHugh et al., Phys. Rev. D {\bf 54}, 5993 (1996).
\bibitem{sm}A. Bodmer, Phys. Rev. D {\bf 4}, 1601 (1971);
         E. Farhi and R. L. Jaffe, Phys. Rev. D {\bf 30}, 2379 (1984). 
\bibitem{Sumiyoshi}K. Sumiyoshi et al., Phys. Rev. D {\bf 42}, 3963 (1990).
\bibitem{evap}K. Sumiyoshi and T. Kajino, Nucl. Phys. B (Proc. Suppl.)
         {\bf 24}, 80 (1991); P. Bhattacharjee et al., Phys. Rev. D {\bf 48},
         4630 (1993).
\bibitem{Hofmann}This mechanism for enhanced baryon penetration through
         the phase interface was pointed out to me by M. Hofmann.
\bibitem{Mathews} In-Saeng Suh and G. J. Mathews, e-print
         astro-ph/9805179 (1998).
\bibitem{perturbations}
         V. Mukhanov and G. Chibisov, Pis'ma Zh. Eksp. Teor. Fiz.
         {\bf 33}, 549 (1981) [JETP Lett. {\bf 33}, 532 (1981)];
         A. Starobinsky, Phys. Lett. B {\bf 117}, 175 (1982);
         A. Guth and S.-Y. Pi, Phys. Rev. Lett. {\bf 49}, 1110 (1982);
         S. Hawking, Phys. Lett. B {\bf 115}, 295 (1982).
\bibitem{Boyd}G. Boyd et al., Phys. Rev. Lett. {\bf 75}, 4169 (1995);
         Nucl. Phys. {\bf B469}, 419 (1996).
\bibitem{MILC97}MILC Collaboration, Phys. Rev. D {\bf 55}, 6861 (1997).
\bibitem{Smoot}G. F. Smoot et al., Astrophys. J. {\bf 396}, L1 (1992).
\bibitem{Sikivie}P. Sikivie, Phys. Rev. Lett. {\bf 51}, 1415 (1983);
         C. Hagmann et al., Phys. Rev. Lett. {\bf 80}, 2043 (1998).
\bibitem{femtolensing}A. Gould, Astrophys. J. {\bf 386}, L5 (1992); 
         A. Ulmer and J. Goodman, Astrophys. J. {\bf 442}, 67 (1995).
\bibitem{microlensing}MACHO Collaboration, Astrophys. J. {\bf 486}, 697 (1997);
         EROS Collaboration, Astron. Astrophys. {\bf 324}, L69 (1997).
\bibitem{GlendenningBook}N. K. Glendenning, {\em Compact Stars} (Springer,
         New York, 1997).
\bibitem{Itho}N. Itho, Prog. Theor. Phys. {\bf 44}, 291 (1970);
         G. Baym and S. Chin, Phys. Lett. {\bf 62B}, 241 (1976);
         G. Chaplin and M. Nauenberg, Nature {\bf 264}, 235 (1976);
         B. D. Keister and L. S. Kisslinger, Phys. Lett. {\bf 64B}, 117 (1976).
\bibitem{Glendenning92}N. K. Glendenning, Phys. Rev. D {\bf 46}, 1274 (1992).
\bibitem{quarkstar}N. K. Glendenning and F. Weber, Astrophys. J. {\bf 400},
         647 (1992); C. Kettner et al., Phys. Rev. D {\bf 51}, 1440 (1995).
\bibitem{Thorsett}S. E. Thorsett, et al., Astrophys. J. {\bf 405}, L29 (1992).
\bibitem{Greiveldinger}C. Greiveldinger et al., Astrophys. J. {\bf 465}, L35
         (1996) and references therein.
\bibitem{Weber}F. Weber, {\em Pulsars as Astrophysical Laboratories for 
         Nuclear and Particle Phsyics}, to be published by IOP Publishing Co,
         Bristol, England.
\bibitem{Boguta}J. Boguta, Phys. Lett. B {\bf 106}, 255 (1981); J. M. Lattimer 
         et al., Phys. Rev. Lett. {\bf 66}, 2701 (1991).
\bibitem{Iwamoto}N. Iwamoto, Phys. Rev. Lett. {\bf 44}, 1637 (1980);
         Ann. Phys. (N.Y.) {\bf 141}, 1 (1982).
\bibitem{BATSE}C. A. Meegan et al., Astrophys. J. Suppl. Ser. 
         {\bf 106}, 65 (1996).
\bibitem{Beppo}E. Costa et al., Nature {\bf 387}, 783 (1997); J. van Paradijs
         et al., Nature {\bf 386}, 686 (1997); D. A. Frail et al., Nature 
         {\bf 389}, 261 (1997).
\bibitem{highz}S. R. Kulkarni et al., Nature {\bf 393}, 35 (1998).
\bibitem{highe}G. J. Fishman et al., Astron. Astrophys. Suppl. Ser. {\bf 97}, 
         17 (1993).
\bibitem{fireball}For two recent reviews see 
         P. M\'esz\'aros, e-print astro-ph/9711354 (1997);
         T. Piran, e-print astro-ph/9801001 (1998).
\bibitem{Eichler}Eichler at al., Nature {\bf 340}, 126 (1989).
\bibitem{Janka}M. Ruffert and H. Th. Janka, e-print astro-ph/9804132 (1998).

\end{thebibliography}
